\documentclass[showpacs,preprintnumbers,amsmath,amssymb]{revtex4}

\usepackage{amsmath}
\usepackage{graphicx}
\usepackage{epsf}
\usepackage{psfrag}
\usepackage{epsfig}
\usepackage{graphics}

\begin{document}

\newcommand{\be}{\begin{equation}}
\newcommand{\ee}{\end{equation}}
\newcommand{\bq}{\begin{eqnarray}}
\newcommand{\eq}{\end{eqnarray}}

\title{{\bf{ Systematization of Basic Divergent Integrals in Perturbation Theory and Renormalization Group Functions}}}

\date{\today}

\author{L. C. T.  Brito$^{(a)}$} \email []{lctbrito@fisica.ufmg.br}
\author{H. G. Fargnoli$^{(a)}$} \email[]{helvecio@fisica.ufmg.br}
\author{A. P. Ba\^eta Scarpelli$^{(b)}$}\email[]{scarp@fisica.ufmg.br}
\author{Marcos Sampaio$^{(a)}$}\email[]{msampaio@fisica.ufmg.br}
\author{M. C. Nemes$^{(a)}$}\email[]{carolina@fisica.ufmg.br}

\affiliation{(a) Federal University of Minas Gerais - Physics
Department - ICEx \\ P.O. BOX 702, 30.161-970, Belo Horizonte MG -
Brazil}
\affiliation{(b) Centro Federal de Educa\c{c}\~ao Tecnol\'ogica - MG \\
Avenida Amazonas, 7675 - 30510-000 - Nova Gameleira - Belo Horizonte
-MG - Brazil}

\begin{abstract}
\noindent
We show that to $n$ loop order the divergent content of a Feynman amplitude is spanned by a set of basic (logarithmically divergent) integrals ${I^{(i)}_{log}(\lambda^2)}$, $i = 1,2, \cdot\cdot\cdot, n$, $\lambda$ being the renormalization 
group scale, which need not be evaluated. Only the coefficients of the basic divergent integrals are show to determine renormalization group functions. Relations between these coefficients of different loop orders are derived. 
\end{abstract}

\pacs{11.10.Gh, 11.15.Bt, 12.38.Bx}

\maketitle

\section{Introduction}

Implicit regularization (IR) is a non-dimensional momentum space  framework which has been claimed to be   a strong candidate for     an invariant regularization suitable to develop perturbation theory in supersymmetric gauge field theories \cite{IR1}-\cite{IR15}. Assuming an implicit regulator in a general (multiloop) Feynman amplitude, a mathematical identity at the level of propagators allows to write the divergent content as basic divergent integrals (BDI) or loop integrals written in terms of one internal momentum only  in an unitarity preserving fashion. This is possible because BPHZ subtractions as well as  the counterterm method are compatible with IR to arbitrary loop order. An arbitrary scale appears via a regularization independent identity which relates two logarithmically BDI' s by trading a mass parameter $m$  (or an infrared regulator in the propagators) for an arbitrary positive parameter $\lambda$, ($[\lambda]=M$) plus a function of $m/\lambda$. Consequently $\lambda$ parametrizes the freedom of separating the divergent content of an amplitude and acts as a renormalization group scale. The key point underlying IR is that neither the (regularization dependent) BDI' s nor their derivatives with respect to $\lambda$ represented by BDI's need be evaluated. In other words, the BDI's are readily absorbed into renormalization constants whose derivatives with respect $\lambda$ used to calculate renormalization group functions can also be expressed by BDI' s.
The advantage of such scheme is that a physical amplitude is written as a finite part plus a set of BDI's say $I_{log}^{(i)} (\lambda^2)$ and finite surface terms (ST's) expressed by volume integrals of a total derivative in momentum space which stem from (finite) differences between $I_{log}^{(i)} (\lambda^2)$ and $I_{log}^{(i) \mu_1 \mu_2 ...} (\lambda^2)$ where the latter is a logarithmically divergent integral which contains in the integrand a product of internal momenta carrying Lorentz indices $\mu_1, \mu_2 ...$. In other words throughout the reduction of the amplitude to loop integrals, $I_{log}^{(i) \mu_1 \mu_2...} (\lambda^2)$ may be written as a product of metric tensors symmetrized in the Lorentz indices times  $I_{log}^{(i)} (\lambda^2)$ plus a surface term. 

Such ST's are in principle arbitrarily valued. However it has  been shown that setting them to zero ab initio corresponds to both invoking translational invariance of Green's functions and allowing  shifts in the integration variable in momentum space \cite{IR4}, \cite{EDSON} which in turn is an essential ingredient to demonstrate gauge invariance based on a diagrammatic proof. Therefore ST's seem to encode the possible symmetry breakings. Moreover it has been verified  that constraining such surface terms to nought  is also sufficient to guarantee that supersymmetry is preserved in the Wess-Zumino model to $3^{nd}$-loop order \cite{IR9} and supergravity to $1$-loop order \cite{IR12}. 
Notwithstanding it is reasonable to assert  that IR is a good candidate to an invariant  calculational friendly  regularization framework valid in arbitrary loop order. From the point of view of algebraic renormalization, ST's would be the necessary symmetry restoring counterterms whose expression is known within IR. Then a constrained version of IR (CIR) amounts to setting them to zero from the start and thus constituting an invariant scheme. 
When physical quantum breakings (anomalies) are expected some care must be exercised: one is able to spot a genuine breaking by letting the ST's to be arbitrary so to verify that none consistent set of values for the ST's dictated by symmetry requirements fulfill all the essential Ward identities of the underlying model at the same time \cite{IR6}, \cite{IR11}. 
 In \cite{IR2}, \cite{EDSON} the rules that define IR  to arbitrary loop order are specified.

A renormalization group equation can immediately be written within IR adopting $\lambda$ as a renormalization group scale and a minimal, mass independent   renormalization scheme in which only the basic divergent integrals are absorbed in the renormalization constants. Hence  renormalized Green's function satisfy a kind of  Callan-Symanzik equation governed by the scale $\lambda$.  

The purpose of this contribution is to twofold. Firstly although
IR works in arbitrary massive quantum field theories, for massless theories it undergoes a remarkable simplification. Assuming an infrared regulator $\mu$ for the propagators, $I_{log}^{(i)} (\mu^2)$ equals $I_{log}^{(i)} (\lambda^2)$, $(\lambda \ne 0)$, plus a sum of terms proportional to powers of the logarithm of the ratio $\mu/\lambda$. We will show in this contribution that for massless theories all the divergencies to arbitrary loop order can be cast as a function of $I_{log}^{(i)} (\lambda^2)$, according to the definition
\begin{equation}
I_{log}^{(i)} (\mu^2) = \int_k^\Lambda \frac{1}{(k^2-\mu^2)^2}\ln^{(i-1)}{\left(-\frac{k^2-\mu^2}{\lambda^2}\right)}.
\label{I_logn}
\end{equation}
where $\int^{\Lambda}_k \equiv \int(d^4k)/(2\pi)^4$ and the superscript $\Lambda$  is a symbol for an implicit 
regularization. Secondly it is well known that  renormalization group functions constitute a testing ground for regularizations because they both encode the symmetry properties of the underlying model which should be preserved by the regularizations and  their expansion  in perturbation theory contains terms which are universal, i. e. renormalization scheme independent. While some interesting simplifications take place in dimensional methods, e.g. in an inverse  power series in $\epsilon \rightarrow 0$ of the coupling constant, beta functions are determined uniquely by the residue of the simple pole on $\epsilon$, it is pertinent to ask what is the counterpart in IR. That is to say, one may wonder how the calculation of renormalization group functions systematizes within a scheme where only basic divergent integrals are claimed to be sufficient to exhibit the ultraviolet properties of a model in a symmetry preserving fashion. The answer to this question is that a general framework for renormalization group functions can be built in which the simplifications of dimensional methods manifest themselves as relations between the coefficients of basic divergent integrals coming from different Feynman graphs that contribute to a given renormalization group function.

We illustrate with the Yukawa model in $3+1$-dimensions  to $2^{nd}$-loop order which contains a $\gamma_5$ matrix and hence the application of dimensional regularization is more involved.

\section{General Ultraviolet Structure of Massless Theories}
\label{General_structure}

The purpose of this section is to show that the ultraviolet content of an amplitude
to $n^{th}$ loop order for massless models, considering the definition,
is written in terms of  $I_{log}^{(i)}(\lambda^2)$.
A  general $n$-loop, $l$-point amplitude, after space-time and internal group algebra contractions are performed, can always  be written as
a combination of integrals of the type
\begin{equation}
\int_{k}^\Lambda  \frac{k_{\mu_1}k_{\mu_2}\cdots k_{\mu_j}}{(k-p_1)^2\cdots (k-p_l)^2}{\cal A}_{n-1}(k,p_1,\cdots,p_l,\lambda^2)\, ,
\label{amp}
\end{equation}
where we have integrated $n-1$ times leaving  only $k$, the most external loop momentum and the $p_i$'s are external momenta.
For a massless model suppose  that ${\cal A}_{n-1}$ is cast like
\be
{\cal A}_{n-1}(k,p_1, \cdots,p_l,\lambda^2) = {\cal A}_{n-1}^\Lambda+ \sum_{i=1}^{n}a_i (k,p_1,\cdots,p_l)
\ln^{i-1}{\left(-\frac{k^2}{\lambda^2}\right)} + \bar{\cal A}_{n-1},
\label{assumption}
\ee
in which $\bar{\cal A}_{n-1}$ is finite under integration on $k$ and
${\cal A}_{n-1}^\Lambda$, the divergent part, represents the subdivergences which in principle are already written in terms of  $I_{log}^{(i)}(\lambda^2)$.
The mass scale $\lambda^2$ has emerged from a scale relations which characterizes a renormalization scheme in Implicit Regularization. The coefficients
$a_i (k,p_1,\cdots,p_l)$ may contain powers in the external and internal momenta.
To justify the assumption of equation (\ref{assumption}) we proceed with a proof by induction. For $n=2$ (one loop order)  it can be easily verified that (\ref{assumption}) holds for ${\cal A}_{1}$ \cite{IR2} . Now we show that this assumption for
$(n-1)^{th}$-loop order implies the same structure for the $n^{th}$-loop order to conclude by induction  
that the multiloop integrals at any order have the same structure.
The relevant contributions  come from the second term on the r.h.s. of (\ref{assumption}),
\begin{equation}
\int_{k}^\Lambda  \frac{k_{\mu_1}\cdots k_{\mu_{r(i)}}}{[(k-p_1)^2-\mu^2]\cdots [(k-p_l)^2-\mu^2]}
\ln^{i-1}{\left(-\frac{k^2-\mu^2}{\lambda^2}\right)},
\end{equation}
which has superficial degree of
divergence $r(i)-2l+4$. Extra factors in the numerator were considered so as to account for the Lorentz structure of the $a_i (k,p_1,\cdots, p_l)$'s. A fictitious mass $\mu^2$ was introduced in the propagators and the limit $\mu^2 \to 0$ will be taken in the end. A fictitious mass may always be introduced if the integral is infrared safe. This is necessary because  although the integral is infrared safe, the expansion of the integrand, as we explain below, breaks  into infrared divergent pieces. When a genuine infrared divergence appears, this procedure can be problematic in non-abelian theories. For such cases a new procedure within IR defining basic infrared divergent integrals is necessary in order to preserve symmetries \cite{WIP}.

 We judiciously apply in the integrand the identity,
\be
\frac {1}{(p_r-k)^2-\mu^2}=\frac{1}{(k^2-\mu^2)} -\frac{p_r^2-2p_r \cdot k}{(k^2-\mu^2) \left[(p_r-k)^2-\mu^2\right]},
\label{ident}
\ee
for the factor in the denominator which depends on $p_l$ to obtain
\bq
&& \sum_{m=1}^{r(i)-2l+5}(-1)^{m-1}\int_{k}^\Lambda
\frac{k_{\mu_1}\cdots k_{\mu_{r(i)}}(p_l^2-2p_l \cdot k)^{m-1}}{(k^2-\mu^2)^m[(k-p_1)^2-\mu^2]\cdots [(k-p_{l-1})^2-\mu^2]}
\ln^{i-1}{\left(-\frac{k^2-\mu^2}{\lambda^2}\right)} \nonumber \\
&& + (-1)^{r(i)-2l+5}
\int_{k}^\Lambda
\frac{k_{\mu_1}\cdots k_{\mu_{r(i)}}(p_l^2-2p_l \cdot k)^{r(i)-2l+5}}{(k^2-\mu^2)^{r(i)-2l+5}[(k-p_1)^2-\mu^2]\cdots [(k-p_l)^2-\mu^2]}
\ln^{i-1}{\left(-\frac{k^2-\mu^2}{\lambda^2}\right)}.
\eq
In the equation above the last integral is finite. We pick out the ultraviolet divergent ones. For each one of the divergent integrals in the summation, the procedure has to be repeated for all the external momenta. Let us consider one typical divergent integral after the expansion has been performed for all the external momentum with the exception of $p_1$:
\be
J_{\mu_1 \cdots \mu_r}= \int_{k}^\Lambda
\frac{k_{\mu_1}\cdots k_{\mu_r}}{(k^2-\mu^2)^\alpha[(k-p_1)^2-\mu^2]}
\ln^{i-1}{\left(-\frac{k^2-\mu^2}{\lambda^2}\right)}.
\label{J}
\ee
For this integral the superficial degree of divergence is $r-2\alpha+2$ and the expansion is performed so as to have the
divergent part freed from the external momentum:
\bq
&&J_{\mu_1 \cdots \mu_r}= \sum_{m=1}^{r-2\alpha+3}(-1)^{m-1}\int_{k}^\Lambda
\frac{k_{\mu_1}\cdots k_{\mu_r}(p_1^2-2p_1 \cdot k)^{m-1}}{(k^2-\mu^2)^{\alpha+m}}
\ln^{i-1}{\left(-\frac{k^2-\mu^2}{\lambda^2}\right)} \nonumber \\
&&+ (-1)^{r-2\alpha+3}
\int_{k}^\Lambda
\frac{k_{\mu_1}\cdots k_{\mu_r}(p_1^2-2p_1 \cdot k)^{r-2\alpha+3}}{(k^2-\mu^2)^{r-\alpha+3}[(k-p_1)^2-\mu^2]}
\ln^{i-1}{\left(-\frac{k^2-\mu^2}{\lambda^2}\right)}.
\label{8}
\eq
For the basic divergent integrals (without dependence on the external momenta), it is only
possible to have an even degree of divergence.
Besides, as shown in ref. \cite{IR2} (from equations $(20)$ to $(23)$ of this reference), it is possible to write a parametrization in which  the quadratic divergences vanish in the limit $\mu^2 \to 0$ to one loop order. The same argument can be generalized to arbitrary loop order. So, we only have to deal with the basic logarithmic
divergent integrals. They have the form,
\begin{equation}
I_{log}^{{(i)}{\mu_1 \cdots \mu_{j}}}(\mu^2)= \int_{k}^\Lambda
\frac{k^{\mu_1}\cdots k^{\mu_{j}}}{(k^2-\mu^2)^p}
\ln^{i-1}{\left(-\frac{k^2-\mu^2}{\lambda^2}\right)},
\end{equation}
where $p=\alpha + m$, $m$ being the summation index in equation (\ref{8}), and $2p= j +  4$, which in turn may always be written in terms of $I_{log}^{(i)}(\mu^2)$'s (see equation (\ref{I_logn})) plus surface terms.
For example, for two Lorentz indices we have
\bq
&&I_{log} ^{(j) \mu \nu} (\mu^2) = \int_k^\Lambda \frac{k^\mu k^\nu}{(k^2-\mu^2)^3}\ln^{j-1}{\left(-\frac{k^2-\mu^2}{\lambda^2}\right)}=
\frac 14 \left\{ g^{\mu \nu}\int_k^\Lambda \frac{1}{(k^2-\mu^2)^2}\ln^{j-1}{\left(-\frac{k^2-\mu^2}{\lambda^2}\right)} \right. \nonumber \\
&& \left.+2(j-1) \int_k^\Lambda \frac{k^\mu k^\nu}{(k^2-\mu^2)^3}\ln^{j-2}{\left(-\frac{k^2-\mu^2}{\lambda^2}\right)}
-\int_k^\Lambda \frac{\partial}{\partial k_\nu}\left[\frac{k^\mu }{(k^2-\mu^2)^2}\ln^{j-1}{\left(-\frac{k^2-\mu^2}{\lambda^2}\right)}\right]
\right\}
\eq
The procedure is repeated for $I_{log}^{(i-1)\mu \nu}$  so to obtain
\be
I_{log} ^{(j) \mu \nu} (\mu^2)= \frac{g^{\mu \nu}}{4} \sum_{i=1}^{j} \frac{1}{2^{j-i}}\frac{(j-1)!}{(i-1)!}
I_{log}^{(i)}(\mu^2)+ \mbox{surface terms}.
\label{surface_terms}
\ee
We still have to deal with the fictitious mass, which in the limit $\mu^2 \to 0$ will give infrared divergent pieces both in the ultraviolet divergent and finite parts. This problem is simply dealt with by the use of regularization independent scale relations
(they can be easily obtained with the help of a cutoff), which read
\begin{equation}
I_{log}^{(j)}(\mu^2)=I_{log}^{(j)}(\lambda^2)-\frac{i}{16\pi^2} \sum_{k=1}^{j} \frac{(j-1)!}{k!}\ln^k{\left(\frac{\mu^2}{\lambda^2}\right)}
\label{scale}
\end{equation}
for arbitrary non-vanishing $\lambda$. This justifies the appearance of the mass scale $\lambda^2$ in ${\cal A}_{n-1}$.
For infrared safe models a systematic cancelation of all powers of $\ln{\left(\frac{\mu^2}{\lambda^2}\right)}$ between
the ultraviolet divergent and finite parts finally crowns $\lambda$ a renormalization group scale.
The important fact here is that an integral of the type (\ref{J}) will have a general result given by
\be
J_{\mu_1 \cdots \mu_r}= \sum_{m=1}^{i}T^{(m)}_{\mu_1 \cdots \mu_r} I_{log}^{(m)}(\lambda^2)
+ \sum_{m=0}^i L^{(m)}_{\mu_1 \cdots \mu_r}\ln^m{ \left(-\frac{p^2}{\lambda^2}\right)},
\ee
where $T^{(m)}_{\mu_1 \cdots \mu_r}$ and $L^{(m)}_{\mu_1 \cdots \mu_r}$ are tensor structures depending on
the external momentum. This permits us to conclude that in the next loop order, the same structure of
divergence will be maintained. The other finite parts, obtained when the expansion of the integrand was carried out,
when inserted into an external loop can give divergent contributions. Nevertheless, they can be put in the form
of (\ref{J}) by performing further expansions. Therefore we can assert that the divergent structure
of massless loop calculations within the context of Implicit Regularization can be completely displayed in terms
of the $I_{log}^{(i)}(\lambda^2)$'s. This completes our proof. 

For the sake of clarity, we exemplify below:
\begin{eqnarray}
&&\int_k^\Lambda \frac{1}{k^2(k - p)^2}\ln^{n-1}\left(-\frac{k^2}{\lambda^2}\right)=
\lim_{\mu^2 \to 0} \left\{ \int_k^\Lambda \frac{1}{(k^2-\mu^2)^2}\ln^{n-1}{\left(-\frac{k^2-\mu^2}{\lambda^2}\right)}
\right. \nonumber\\ && \left. -
\int_k\frac{p^2-2p \cdot k}{(k^2-\mu^2)^2[(k - p)^2-\mu^2]}\ln^{n-1}\left(-\frac{k^2-\mu ^2}{\lambda^2}\right) \right\},
\end{eqnarray}
which for $n=3$ yields
\begin{eqnarray}
&&  \lim_{\mu^2 \to 0} \left\{I_{log}^{(3)}(\mu^2)+ \frac{i}{16\pi^2} \sum_{k=1}^{3}
\frac{2!}{k!}\ln^k{\left(\frac{\mu^2}{\lambda^2}\right)}
+ \frac{i}{16\pi^2}
 \left[ 2 - \sum_{k=0}^{3}(-1)^{3-k} \frac{2!}{k!}\ln^k{\left(-\frac{p^2}{\lambda^2}\right)}\right] \right\}
\nonumber \\
&& =I_{log}^{(3)}(\lambda^2) + \frac{i}{8\pi^2} \left\{ 1 - \sum_{k=0}^{3}(-1)^{3-k} \frac{1}{k!}
\ln^k{\left(-\frac{p^2}{\lambda^2}\right)}\right\},
\end{eqnarray}
where in the last step we have used (\ref{scale}) for $j=3$.

\section{Renormalization Group Functions}

In this section we present a general framework to work out renormalization group functions using renormalization constants defined by BDI's. We will see that  derivatives of BDI's which are  also BDI's need not be evaluated. Moreover, in the calculation of renormalization group functions, a simplification becomes manifest through relations between some coefficients of BDI's. We study the massless Yukawa theory in $3+1$-dimensions to $2$-loop order as a working example because both  it is rich enough due to the presence of overlapping divergences and two coupling constants.   Besides, dimensional methods  are more involved as a $\gamma_5$ matrix appears in the interaction term.

The Lagrangian density in terms of renormalized variables $\phi_{0} = Z_{\phi}^{\frac{1}{2}}\phi$, $\psi_{0}=Z_{\psi}^{\frac{1}{2}}\psi$, $e_{0}= e Z_{e}/(Z_{\psi}Z_{\phi}^{\frac{1}{2}})$, $g_{0}=g  Z_g/Z_{\phi}^{2}$ reads 
\begin{equation}
\mathcal{L} = (1+A) \partial_{\mu}\phi\partial\phi^{\mu} +  i (1+B) \bar{\psi}\gamma_{\mu}\partial^{\mu}\psi
+ i(1+C)  e \bar{\psi}\gamma^{5}\psi\phi- (1+D) \frac{g}{4!}\phi^{4},
\label{yukawua}
\end{equation}
where $Z_{\phi} = 1 + A$, $Z_{\psi} = 1 + B$, $Z_{e} = 1 + C$ e $Z_{g} = 1 + D$.

\begin{figure}
\epsffile{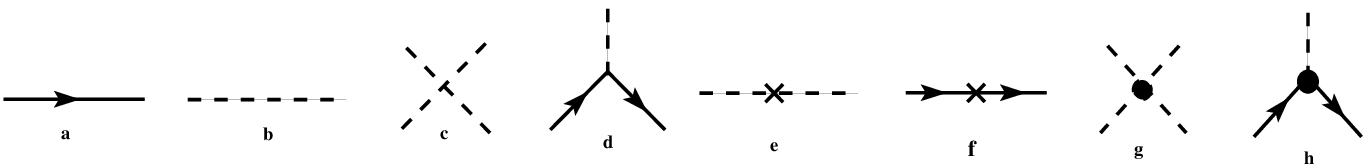}
\caption{Feynman rules to the Yukawa model: $(\mbox{a})\to i /\not\!p$, $(\mbox{b})\to i/p^2$, $(\mbox{c}) \to -i g$, $(\mbox{d}) \to - e \gamma_5$, $(\mbox{e}) \to i A p^2$, $(\mbox{f}) \to i B\not\!p $, $(\mbox{g}) \to -i D g$, $(\mbox{h}) \to - e \gamma_5 C$}
\label{feynman_rules}
\end{figure}
The superficial degree of divergence for any graph with $n_{\phi}$ external boson lines and $n_{\psi}$ external fermion lines is given by $ \Delta = 4 - n_{\phi} - \frac{3}{2} n_{\psi} $ thus to $2$ loop order the divergent amplitudes are  $\Gamma_{\phi^2}$ ($n_{\phi} = 2$ and $n_{\psi}$ = 0), $\Gamma_{\bar{\psi}\psi}$ ($n_{\phi} = 0$ and $n_{\psi} = 2$ ), $\Gamma_{\bar{\psi}\psi\phi}$  ($n_{\phi} = 1$ e $n_{\psi} = 2$ ) and  $\Gamma_{\phi^4}$ ($n_{\phi} = 4$ and $n_{\psi} = 0$ ) .  
Next we evaluate the diagrams necessary to compute the renormalization group functions to $2$-loop order, which are portraited
in figures $2$ to $5$. It is not difficult to show that the amplitudes following from the Feynman rules can be treated in IR by separating the external momentum dependence in the BDI's using  (\ref{ident}) and neglecting surface terms which stem from (\ref{surface_terms}) whilst the renormalization group scale is defined through (\ref{scale}). We summarize the results below.
In figure $2$ the divergences can be isolated as
\begin{eqnarray}
\Gamma_{{\rm 2a}} &=& i A p^2\\
\Gamma_{{\rm 2b}} &=& 2 e^2 p^2 I_{log}(\lambda^2), \\
\Gamma_{{\rm 2c}} + \Gamma_{{\rm 2d}} &=& \frac{e^4 p^2}{4\pi^2} \left[2 I_{log}(\lambda^2) - i 16 \pi^2 \left[ I_{log}(\lambda^2)\right]^{2}- I_{log}(\lambda^2)\ln\left(-\frac{p^2}{\lambda^2}\right) \right], \label{cd}\\
\Gamma_{{\rm 2e}} &=& \frac{e^4 p^2 }{16 \pi^2} \left[ 2  I_{log}(\lambda^2) - i 16 \pi^2 \left[ I_{log}(\lambda^2)\right]^{2} -  I_{log}(\lambda^2)\ln\left(-\frac{p^2}{\lambda^2}\right)  \right] , \label{e}\\
\Gamma_{{\rm 2f}} &=&  \frac{e^{4}p^{2}}{16\pi^{2}}\left[ - \frac{9}{2}I_{log}(\lambda^{2}) + 16\pi^{2}i\left[I_{log}(\lambda^{2})\right]^{2} + \ln\left(-\frac{p^{2}}{\lambda^{2}}\right)I_{log}(\lambda^{2}) + I_{log}^{(2)}(\lambda^{2})\right],\\
\Gamma_{{\rm 2g}} &=&  \frac{e^4 p^2}{8\pi^2}   \left[- 5  I_{log}(\lambda^2) + i 16 \pi^2 \left[ I_{log}(\lambda^2)\right]^{2} + 2  I_{log}(\lambda^2)\ln\left(-\frac{p^2}{\lambda^2}\right)\right] ,\\
\Gamma_{{\rm 2h}} &=& \frac{g^2 p^2}{12 (4\pi)^2}  I_{log}(\lambda^2),  \\
\end{eqnarray}

and  $\Gamma_{\phi^2} =  i A p^2 + \Gamma_{2{\rm b}} + \Gamma_{2{\rm c}} + \Gamma_{2{\rm d}} + \Gamma_{2{\rm e}} +   \Gamma_{2{\rm f}} + \Gamma_{2{\rm g}} + \Gamma_{2{\rm h}}$ which yields
\be
\Gamma_{\phi^2} = i A p^2 + \frac{p^2}{16\pi^2}\left[\left( 32\pi^2 e^2 -  7e^4 + \frac{g^2}{12} \right) I_{log}(\lambda^2) - 32\pi^2 i e^4 \left[I_{log}(\lambda^2)\right]^{2} +  2 e^4 I^{(2)}_{log}(\lambda^2)\right].
\label{i}
\ee
In determining the counterterm graphs that correspond to the amplitudes in equations (\ref{cd}) and (\ref{e}) above we have used the one loop contributions of the counterterms $B$ and $C$ from equation (\ref{contratermos}). Notice that
the non-local divergences have been correctly canceled as they should because we have shown in \cite{IR2} that IR is compatible with the counterterm method derived from BPHZ forest formula.
\begin{figure}
\includegraphics{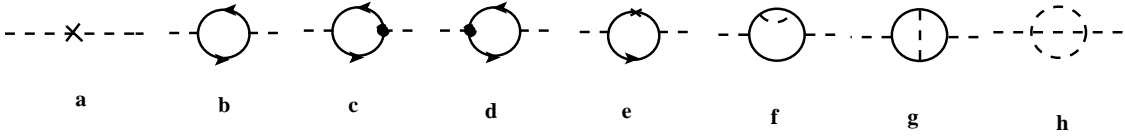}
\caption{Diagrams contributing to $\Gamma_{\phi^2}$.}
\label{phi2}
\end{figure}
Diagrams in figures $3$, $4$ and $5$ are evaluated in a similar fashion, using one loop counterterms previously determined, to give, 
\be
\Gamma_{\bar{\psi}\psi} = iB\not\!p + \frac{\not\!p}{16\pi^{2}}\left[\left(8\pi^{2}e^{2}-\frac{31}{8}e^{4}\right)I_{log}(\lambda^{2})+\frac{9}{4}e^{4}I_{log}^{(2)}(\lambda^{2})\right],
\label{ii}
\ee

\begin{figure}
\includegraphics{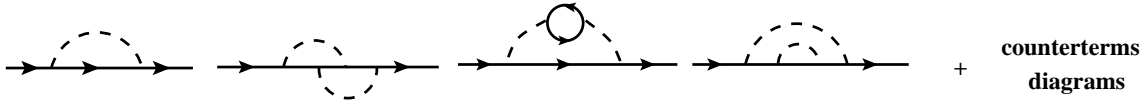}
\caption{Diagrams contributing to $\Gamma_{\bar{\psi}\psi}$.}
\label{psi2}
\end{figure}

\begin{equation}
\Gamma_{\bar{\psi}\psi\phi} = -i e C\gamma_{5} + \frac{i\gamma_{5}}{16\pi^{2}} \left[\left(-16\pi^{2}e^{3} + 
g e^{3} + 9e^{5} \right)I_{log}(\lambda^{2})-6e^{5}I_{log}^{(2)}(\lambda^{2})\right],
\label{iii}
\end{equation}

\begin{figure}
\includegraphics{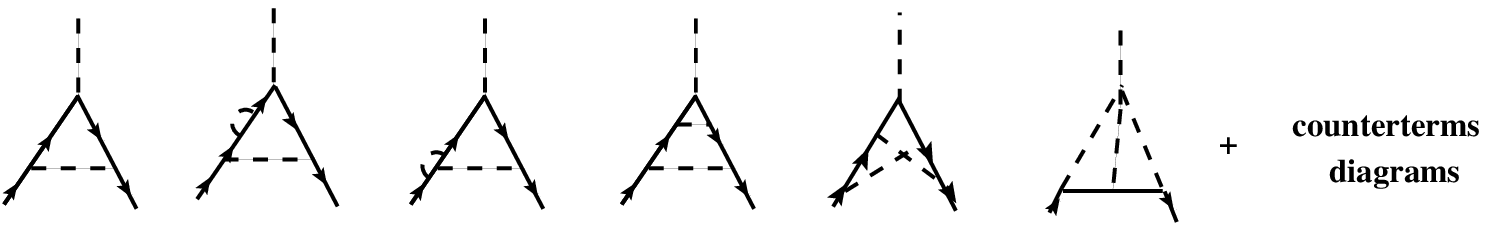}
\caption{Diagrams contributing to $\Gamma_{\bar{\psi}\psi\phi}$.}
\label{phipsipsi}
\end{figure}

\begin{eqnarray}
\Gamma_{\phi^4} &=&  -i g D + \left(24i\pi^{2}g^{2}-384\pi^{2}e^{4}-6 e^{3} -12 g^{2}e^{2}+336e^{6} + 96 g e^{4}\right)\frac{I_{log}(\lambda^{2})}{16\pi^{2}}
\nonumber\\
&+& \left(3 g^{3}+6g^{2}e^{2}-144e^{6} -72 g e^{4}\right)\frac{I_{log}^{(2)}(\lambda^{2})}{16\pi^{2}}+ i \left(-\frac{3}{4}g^{3}-96e^{6}-72 g e^{4}\right)[I_{log}(\lambda^{2})]^{2},
\label{iv}
\end{eqnarray}
respectively.

\begin{figure}
\includegraphics{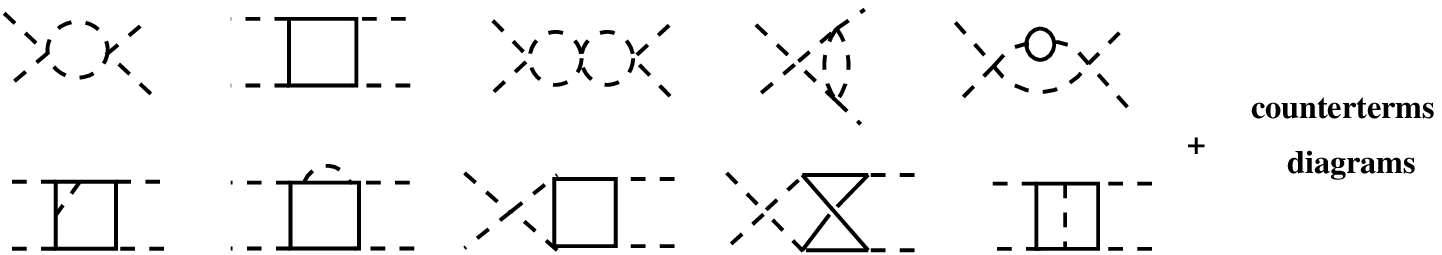}
\caption{Diagrams contributing to $\Gamma_{\phi^4}$.}
\label{phi4}
\end{figure}
Then we have the renormalization constants defined in a minimal scheme if
\begin{eqnarray}
A &=& \frac{i}{16\pi^{2}}\left[\left(\frac{g^{2}}{12}+32\pi^{2}e^{2}\right)I_{log}(\lambda^{2})+e^{4}\left(-7I_{log}(\lambda^{2})-32i\pi^{2}\left[I_{log}(\lambda^{2})\right]^{2}+2I_{log}^{(2)}(\lambda^{2})\right)\right],\nonumber \\
B &=& \frac{i}{16\pi^{2}}\left[e^{2}8\pi^{2}I_{log}(\lambda^{2})+e^{4}\left(-\frac{31}{8}I_{log}(\lambda^{2})+\frac{9}{4}I_{log}^{(2)}(\lambda^{2})\right)\right], \nonumber \\ 
C &=& \frac{1}{16\pi^{2}}\left[-e^{2}16\pi^{2}I_{log}(\lambda^{2})+e^{2} g I_{log}(\lambda^{2})+e^{4}\left(9I_{log}(\lambda^{2})-6I_{log}^{(2)}(\lambda^{2})\right)\right], \nonumber \\
D &=& \frac{1}{16\pi^{2}}[ g 24\pi^{2}I_{log}(\lambda^{2})+ g^{2}\left(6iI_{log}(\lambda^{2})-3iI_{log}^{(2)}(\lambda^{2})-12\pi^{2}[I_{log}(\lambda^{2})]^{2}\right)\nonumber \\
&+&g^{-1}e^{4}384i\pi^{2}I_{log}(\lambda^{2})+ g^{-1}e^{6}\left(-336iI_{log}(\lambda^{2})+144iI_{log}^{(2)}(\lambda^{2})-96(16\pi^{2})[I_{log}(\lambda^{2})]^{2}\right)\nonumber\\
&+& e^{4}\left(-96iI_{log}(\lambda^{2})+72iI_{log}^{(2)}(\lambda^{2})-72(16\pi^{2})[I_{log}(\lambda^{2})]^{2}\right)]\label{D},
\label{contratermos} 
\end{eqnarray} 

The beta-functions and field anomalous dimensions are defined as usual
\begin{eqnarray}
\gamma_{\phi} &=& \frac{\lambda^2}{Z_{\phi}}  \frac{\partial Z_{\phi}}{\partial\lambda^{2}},
\gamma_{\psi} = \frac{\lambda^2}{Z_{\psi}}  \frac{\partial Z_{\psi}}{\partial\lambda^{2}}, \nonumber\\
\beta_{e} &=& - g\lambda^{2}\left(Z_{e}^{-1}\frac{\partial Z_{e}}{\partial\lambda^{2}}-\frac{1}{2}Z_{\phi}^{-1}\frac{\partial Z_{\phi}}{\partial\lambda^{2}}-Z_{\psi}^{-1}\frac{\partial Z_{\psi}}{\partial\lambda^{2}}\right), 
\beta_{g} = - 2 g \lambda^{2}\left(Z_{g}^{-1}\frac{\partial Z_{g}}{\partial\lambda^{2}}-2Z_{\phi}^{-1}\frac{\partial Z_{\phi}}{\partial\lambda^{2}}\right),
\label{renorm_group}
\end{eqnarray}
where $\lambda$ is the IR arbitrary scale which plays the role of renormalization group scale. To $n$-loop order, a general 
renormalization constant can be written as
\begin{equation}
Z =1+\sum_{j=1}^{n} \Big( g ^{p}Z_{g}^{(j)}+ e^{q}Z_{e}^{(j)}+ g^{r}e^{s}Z_{ge}^{(j)} \Big).
\label{general}
\end{equation}
in which $p, q, r, s$ assume positive integer values in each $n$. In a minimal, mass independent renormalization scheme, 
the renormalization constants $Z_{g}^{(j)}$, $Z_{e}^{(j)}$ and  $Z_{ge}^{(j)}$  take the general form
\begin{equation}
Z^{(j)} = \sum_{k=1}^{j}A_{k}^{(j)} [I_{log}(\lambda^{2})]^{k}+ \sum_{k=2}^{j}  B_{k}^{(j)}
I_{log}^{(k)}(\lambda^{2}) .
\label{Z_(n)}
\end{equation}
For the Yukawa model to two loop order we have, 
\begin{eqnarray}
Z_{\phi} &=& 1 +  g^{2} Z^{(2)}_{a} + \sum_{n=1}^{2} e^{2n} Z^{(n)}_{b}\label{Z_phi2loop}\nonumber\\
Z_{\psi} &=& 1 + \sum_{n=1}^{2}Z^{(n)}_{c} e^{2n},\nonumber \\
Z_{e} &=& 1 + e^{2} g Z^{(2)}_{d} + \sum_{n=1}^{2}Z^{(n)}_{e} e^{2n}, \nonumber\\
Z_{g} &=& 1 + e^{4}Z^{(2)}_{f}+ g e^{2}Z^{(2)}_{g} + \sum_{n=1}^{2}\left(g^{-1}e^{2n+2}Z^{(n)}_{h}+ g^{n}Z^{(n)}_{i}\right).
\label{Z_n} 
\end{eqnarray}
Now plugging equations (\ref{Z_n}) into (\ref{renorm_group}) permits us to obtain the finite contributions to renormalization group functions to $1$ and $2$-loop order from
\begin{eqnarray}
\gamma_{\phi}^{(1)} &=& e^2 \lambda^2 \frac{\partial Z^{(1)}_{b}}{\partial\lambda^{2}}, \,\,\gamma_{\psi}^{(1)} =  e^ 2\lambda^{2} \frac{\partial Z_{c}^{(1)}}{\partial\lambda^{2}}, \nonumber \\
\beta_{e}^{(1)}&=& 2 e^{3}\lambda^{2}\left(\frac{1}{2}\frac{\partial Z^{(1)}_{b}}{\partial\lambda^{2}}-\frac{\partial Z^{(1)}_{e}}{\partial\lambda^{2}}+\frac{\partial Z^{(1)}_{c}}{\partial\lambda^{2}}\right), \nonumber \\
\beta_{g}^{(1)}&=&4 g e^{2}\lambda^{2}\frac{\partial Z^{(1)}_{b}}{\partial\lambda^{2}}- 2 e^{4}\lambda^{2}\frac{\partial Z^{(1)}_{h}}{\partial\lambda^{2}}-  2 g^ 2 \lambda^2 \frac{\partial Z^{(1)}_{i}}{\partial\lambda^{2}},
\label{1loop}
\end{eqnarray}
and 
\begin{eqnarray}
\gamma_{\phi}^{(2)} &=& \lambda^2 \left(g^{2}\frac{\partial Z^{(2)}_{a}}{\partial\lambda^{2}}+e^{4}\frac{\partial Z^{(2)}_{b}}{\partial\lambda^{2}}\right), \,\, \gamma_{\psi}^{(2)} = e^4 \lambda^{2} \frac{\partial Z_{c}^{(2)}}{\partial\lambda^{2}}, \nonumber \\
\beta_{e}^{(2)}&=&- 2 e^{3} g \lambda^{2}\frac{\partial Z^{(2)}_{d}}{\partial\lambda^{2}}+ e g^{2}\lambda^{2}\frac{\partial Z^{(2)}_{a}}{\partial\lambda^{2}}+ 2 e^{5}\lambda^{2}\left(\frac{\partial Z^{(2)}_{c}}{\partial\lambda^{2}}+\frac{1}{2}\frac{\partial Z^{(2)}_{b}}{\partial\lambda^{2}}-\frac{\partial Z^{(2)}_{e}}{\partial\lambda^{2}}\right),\nonumber \\
\beta_{g}^{(2)} &=& 2 g e^{4}\lambda^{2}\left(\frac{\partial Z^{(2)}_{b}}{\partial\lambda^{2}}-  \frac{\partial Z^{(2)}_{f}}{\partial\lambda^{2}}\right)+ 2 g^{3}\lambda^{2}\left( \frac{\partial Z^{(2)}_{a}}{\partial\lambda^{2}}- \frac{\partial Z^{(2)}_{i}}{\partial\lambda^{2}}\right) \nonumber \\
&-& 2 e^{6}\lambda^{2}\frac{\partial Z^{(2)}_{h}}{\partial\lambda^{2}}- 2 e^{2}g^{2}\lambda^{2}\frac{\partial Z^{(2)}_{g}}{\partial\lambda^{2}}.
\label{2loop}
\end{eqnarray}
To complete our task we have to evaluate the derivatives of (\ref{Z_(n)}) w.r.t $\lambda^2$ which are expressible in terms of BDI's as well, namely
\be
\lambda^ 2 \frac{\partial Z^{(n)}_\alpha}{\partial \lambda^2} = -\frac{i}{16 \pi^{2}}\Bigg[A_{\alpha 1}^{(n)}+\sum_{j=2}^{n}(j-1)!B_{\alpha j}^{(n)}
+ \sum_{k=2}^{n} \Bigg( kA_{\alpha k}^{(n)}\left[I_{log}(\lambda^{2})\right]^{k-1}-16i\pi^{2}(k-1)B_{\alpha k}^{(n)}I_{log}^{(k-1)}(\lambda^{2})\Bigg)\Bigg],
\label{deriv_nloop}
\ee
$\alpha=a, \cdots i$, which is a general expression for massless models though a similar one holds for massive models as well. Direct inspection of equation (\ref{contratermos}) enables us to determine the coefficients $A_{\alpha 1}^{(n)}$ and $B_{\alpha j}^{(n)}$ which appear in (\ref{1loop}) and (\ref{2loop}). After some straightforward algebra we obtain 
\be
\gamma^{(1)}_{\phi} =  2 \frac{e^{2}}{(4\pi)^{2}} ,\,\, 
\gamma^{(1)}_{\psi} = \frac{1}{2}\frac{e^{2}}{(4\pi)^{2}} , \,\,
\beta^{(1)}_{e} = 5\frac{e^{3}}{(4\pi)^{2}} , \,\, 
\beta^{(1)}_{g} = 3\frac{g^{2}}{(4\pi)^{2}} - 48\frac{e^{4}}{(4\pi)^{2}} + 8\frac{g e^{2}}{(4\pi)^{2}}
 \label{GammaBetaoneloop}
\ee
and
\begin{eqnarray}
&&\gamma^{(2)}_{\phi} =  \frac{1}{12}\frac{g^{2}}{(4\pi)^{4}} - 5\frac{e^{4}}{(4\pi)^{4}}, \,\, \gamma^{(2)}_{\psi} = - \frac{13}{8}\frac{e^{4}}{(4\pi)^{4}}, \nonumber \\
&& \beta^{(2)}_{e} = -\frac{57}{4}\frac{e^{5}}{(4\pi)^{4}} + \frac{1}{12}\frac{g^{2} e}{(4\pi)^{4}} - 2\frac{g e^{3}}{(4\pi)^{4}}, \,\, 
\beta^{(2)}_{g} = -\frac{17}{3}\frac{g^{3}}{(4\pi)^{4}} + 384\frac{e^{6}}{(4\pi)^{4}} - 12\frac{g^{2}e^{2}}{(4\pi)^{4}} + 28\frac{g e^{4}}{(4\pi)^{4}},
\label{GammaBetatwoloop}
\end{eqnarray}
which agree with \cite{schubert}, \cite{cristina}.

We can generalize (\ref{1loop}) and (\ref{2loop}) to arbitrary loop order using the expansion  (\ref{general})  in (\ref{renorm_group}) to conclude that all we need to evaluate the renormalization functions is the derivative of $Z^{(n)}$
as given in (\ref{deriv_nloop}).
It is interesting to remark that whilst the finite terms in the r.h.s of (\ref{deriv_nloop}) contribute to the computation of the renormalization group 
functions, the terms proportional to BDI's will give relations between   $A_{k}^{(n)}$ and $B_{k}^{(n)}$ as they must vanish because the renormalizations functions are finite. The same reasoning leads us to conclude, in dimensional regularization methods, that only residues of order one contribute to beta-functions. For instance, from  the calculation of the field anomalous dimensions $\gamma_\phi$ and $\gamma_\psi$ up to two loop order  we get,  
\begin{eqnarray}
A_{a2}^{(2)}-8i\pi^{2}B_{a2}^{(2)} &=& 0 \label{cond1finite}\\
i\left(A_{b2}^{(2)}-8i\pi^{2}B_{b2}^{(2)}\right)&=&\frac{1}{2}\left(A_{b1}^{(1)}\right)^{2}+A_{b1}^{(1)}
\left(-\frac{1}{2}A_{b1}^{(1)}+A_{e1}^{(1)}-A_{c1}^{(1)}\label{cond2finite}\right).
\end{eqnarray}
and
\begin{equation}
i\left(A_{c2}^{(2)}-8i\pi^{2}B_{c2}^{(2)}\right)=\frac{1}{2}\left(A_{c1}^{(1)}\right)^{2}+A_{c1}^{(1)}\left(-\frac{1}{2}A_{b1}^{(1)}+A_{e1}^{(1)}-A_{c1}^{(1)}\right)
\label{cond3finite}
\end{equation}
respectively.

To conclude we have shown that in Implicit Regularization (IR), we can organize the divergent content of an amplitude 
to $n^{th}$ loop order in terms of a basis of basic divergent integrals (BDI's), namely $\{ I_{log}^{(i)}(\lambda^2)\}$, $i=1\cdots n$ where $\lambda$ is the RG scale. The calculation of RG functions systematizes within IR for they can be written in terms of coefficients of BDI's. Such coefficients are shown to be inter-related which in turn allows us to restrict
ourselves to a subset of BDI's at each loop order  to evaluate RG functions.

\section*{Acknowledgements}

This work was supported by the Brazilian agency Conselho Nacional de Desenvolvimento Cient\'{\i}fico e 
Tecnol\'{o}gico (CNPq). 
\section*{Appendix}

We calculate explicit the diagram g of figure 2:

\begin{equation}
\Gamma_{2g} = -\mbox{tr}\int^{\Lambda}_{k}\int^{\Lambda}_{l}(-g\gamma_{5})\frac{i}{\not\!l}(-g\gamma_{5})\frac{i}{\not\!k}(-g\gamma_{5})\frac{i}{\not\!k - \not\!p}(-g\gamma_{5})\frac{i}{\not\!l - \not\!p}\frac{i}{(k-l)^{2}}
\end{equation}
where $l$ and $k$ are internal momenta. Taking the trace of Dirac matrices and simplifing we obtain:

\begin{equation}
\Gamma_{2g} = -2ig^{4}\int^{\Lambda}_{k}\int^{\Lambda}_{l}\frac{k^{2}(p-l)^{2} + l^{2}(p-k)^{2}-p^{2}(l-k)^{2}}{l^{2}k^{2}(k-p)^{2}(l-p)^{2}(k-l)^{2}}
\end{equation}
or
\begin{equation}
\Gamma_{2g} = -2ig^{4}\left\{-p^{2}\int_{k}^{\Lambda}\frac{1}{k^{2}(k-p)^{2}}\int_{l}^{\Lambda}\frac{1}{l^{2}(l-p)^{2}}+
2\int_{k}^{\Lambda}\frac{1}{k^{2}}\int_{l}^{\Lambda}\frac{1}{(l-p)^{2}(k-l)^{2}} \right\}
\end{equation}

At this point we apply in each of these integrals the methods discussed in section II. After some algebra we get:

\begin{equation}\label{diagparcial}
\Gamma_{2g} = \frac{g^{4}p^{2}}{8\pi^{2}}\left(-5I_{log}(\lambda^{2})+i16\pi^{2}\left[I_{log}(\lambda^{2})\right]^{2} + 2I_{log}(\lambda^{2})\ln \left(-\frac{p^{2}}{\lambda^{2}}\right) + \mbox{finite}\right)
\end{equation}

Observe that the third term on the r.h.s. of (\ref{diagparcial}) is non-local and it must be canceled with the ones of the counterterm diagrams.

\end{document}